 \documentclass{article} 

\usepackage{lineno,hyperref}
\modulolinenumbers[5]

\usepackage{cite}

\usepackage{amsmath,amssymb}
\usepackage{graphicx}
\usepackage{dcolumn}
\usepackage{booktabs}
\usepackage{multirow}
\usepackage{bm}
\usepackage{hyperref}
\usepackage{braket}
\usepackage{soul}

\usepackage{xcolor}						

\begin{document}

\begin{center}

{\Large \bf 
A bound on energy extraction \\ (and hairiness)  from superradiance
}
\vspace{0.8cm}
\\
{Carlos A. R. Herdeiro$^{\dagger}$, Eugen Radu$^{\dagger}$ 
and  
Nuno M. Santos$^{\dagger, \star}$
\\
\vspace{0.3cm}
$^{\dagger }${\small Departamento de Matem\'atica da Universidade de Aveiro and } \\ {\small  Centre for Research and Development  in Mathematics and Applications (CIDMA),} \\ {\small    Campus de Santiago, 3810-183 Aveiro, Portugal}
\\
\vspace{0.3cm}
$^{\star}${\small Centro de Astrof\'{\i}sica e Gravita\c c\~ao  -- CENTRA, Departamento de F\'{\i}sica, \\
Instituto Superior T\'ecnico -- IST, Universidade de Lisboa -- UL,}
\\ {\small Avenida
Rovisco Pais 1, 1049-001 Lisboa, Portugal}
\vspace{0.3cm}
}
\end{center}
 
\begin{abstract}
The possibility of mining the rotational energy from black holes has far--reaching implications. Such  energy extraction could occur even for isolated black holes, if hypothetical ultralight bosonic particles exist in Nature, leading to a new equilibrium state -- a black hole with synchronised bosonic hair -- whose lifetime could exceed the age of the Universe. A natural question is then: for an isolated black hole and at maximal efficiency, how large is the energy fraction $\epsilon$ that can be extracted from a Kerr black hole by the superradiant growth of the dominant mode? In other words, how hairy can the resulting black hole become? A thermodynamical bound for the total superradiance efficiency, $\epsilon\lesssim 0.29$ (as a fraction of the initial black hole mass), has long been known, from the area law. However, numerical simulations exhibiting the growth of the dominant mode only reached about one third of this  value. We show that if the development of superradiant instabilities is approximately conservative (as suggest by the numerical evolutions), this efficiency is limited to  $\epsilon\lesssim 0.10$, regardless of the spin of the bosonic field. This is in agreement with the maximum energy extraction obtained in numerical simulations for a vector field and predicts the result of similar simulations with a scalar field, yet to be performed.  
\end{abstract}




\section{\label{sec:level1}Introduction}

The recent successes of radio~\cite{Akiyama:2019cqa} and gravitational--wave astronomy\cite{Abbott:2016blz,LIGOScientific:2018mvr,LIGOScientific:2020ibl} have put general relativity (GR) to the test as never before. A key issue is the observational confirmation of the \textit{Kerr hypothesis}: that all astrophysical black holes (BHs), regardless of their scale, are well described by Kerr's solution to the vacuum Einstein's field equations~\cite{Kerr:1963ud}. From a theoretical standpoint, this paradigm relies on: $(i)$ the \textit{uniqueness theorems} of vacuum GR \cite{Carter:1971zc,Robinson:1975bv,Chrusciel:2012jk}, which establish that its most general solution, regular on and outside the event horizon, is the Kerr metric, solely defined by its global charges, mass $M$ and angular momentum $J$; $(ii)$ a number of \textit{no--hair theorems} ($e.g.$~\cite{Bekenstein:1972ny,Herdeiro:2015waa}) ruling out the existence of non--Kerr BHs in the presence of certain types of matter--energy, which otherwise could endow BHs with ``hair". Consequently, these theorems support the \textit{no--hair conjecture} \cite{Ruffini:1971bza}, according to which the gravitational collapse of any type of matter--energy in GR always yields a Kerr BH. 

At the time of writing there is no clear tension between the Kerr hypothesis and observations. Yet, fundamental open issues such as dark matter, dark energy and the inevitability of singularities in GR\cite{Penrose:1964wq,Hawking:1970zqf} strongly suggest going beyond GR and/or the standard model of particle physics. In this context, dynamically robust non--Kerr models are particularly welcome as exploratory scenarios of deviations from the Kerr hypothesis. Clearly, a formation scenario and sufficient stability are mandatory to make any alternative BH (or exotic compact object) model physically plausible and a potential actor on the astrophysical stage.

In this discussion, the phenomenon of superradiance~\cite{Brito:2015oca} originates a novel possibility. Bosonic fields with a mass in an appropriate range can efficiently transfer the rotational energy of a Kerr BH into a cloud of bosonic particles, spinning down the BH -- see $e.g.$~\cite{Arvanitaki:2010sy}. When the BH spins down enough to meet the phase angular velocity of the dominant superradiant mode, the process stalls. The detailed phenomenology depends on the type of bosonic field. Real fields have a rich phenomenology related to the decay of the bosonic cloud via gravitational waves emission -- see $e.g.$~\cite{Arvanitaki:2010sy,Ng:2020ruv,Yuan:2021ebu}. If the (scalar or vector) field is complex, on the other hand, the cloud is stationary after the synchronisation between the BH's and the  field's angular velocities occurs, and a new stationary equilibrium state forms. This has been seen in the  numerical evolutions of East and Pretorius, focusing on the case of vector bosonic fields~\cite{East:2017ovw}. The new equilibrium states were shown to match  Kerr BHs with synchronised Proca hair~\cite{Herdeiro:2017phl}, first reported in~\cite{Herdeiro:2016tmi} following the construction of BH solutions with synchronised scalar hair~\cite{Herdeiro:2014goa}. 

The BHs with synchronised hair (BHsSH) formed from the superradiant instability of Kerr are, themselves, prone to superradiant instabilities~\cite{Ganchev:2017uuo}. These, however, can be very long--lived; in particular, for supermassive BHs the lifetime of the superradiant instability of the newly formed hairy BH can take a timescale larger than a Hubble time to develop~\cite{Degollado:2018ypf}. In other words, \textit{superradiance can form a hairy BH on an astrophysical timescale, and the latter can be stable for a cosmological timescale}. This makes these hairy BHs resulting from the growth of the dominant superradiant mode of Kerr plausible players in astrophysical processes.

The superradiance scenario offers a formation mechanism for BHsSH with an interesting twist. Superradiance is quite sensitive to a matching of scales. The instability is strongest if the Compton wavelength of the bosonic particle, $1/\mu$, and the BH radius, $\sim M$, approximately match: $M\mu~\sim 1$.  Away from this sweet spot, the timescale of the instability grows with the exponential of $M\mu$  for $M\mu\gg 1$~\cite{Zouros:1979iw} and as a large inverse power of $M\mu$ for $M\mu\ll 1$~\cite{Detweiler:1980uk}. Thus, BHs only become efficiently hairy (on an astrophysical timescale) in an island of the parameter space, determined by the mass of the bosonic particle $\mu$. Sufficiently far from the resonance $M\mu~\sim 1$, Kerr BHs are effectively stable against superradiance. Thus, in this scenario, Kerr and hairy BHs may co--exist in the Universe, with the latter belonging to an island in a narrow mass range around $\mu$ (and non--zero spin) and Kerr BHs composing the surrounding ocean of the BH mass spectrum. 

 For astrophysical BHs, known to exist within the mass range $M\in [1,10^{10}] M_\odot$, the resonance $M\mu\sim 1$ means that the bosonic particle is ultralight, with the mass range $\mu\sim [10^{-10},10^{-20}]$ eV. This connects to the particular class of dark matter models known as fuzzy dark matter~\cite{Suarez:2013iw,Hui:2016ltb}, which could have a stringy origin~\cite{Arvanitaki:2009fg}, or otherwise be embedded in simpler extensions of the standard model -- see $e.g.$~\cite{Freitas:2021cfi}.

The foregoing discussion yields the exciting possibility that BHs (and in this case the hairiness of BHs) can become particle detectors of extremely light -- and potentially  inaccessible to colliders -- dark matter particles, via astrophysical measurements. To assess this possibility, however, it is important to understand \textit{how much energy} can be extracted from a Kerr BH from the growth of the dominant superradiant mode. Or, in the context of the complex bosons that endow BHs with hair, this translates into \textit{how hairy} a BH can become, which naturally defines how non--Kerr its phenomenology may be. In his pioneering paper on the area law, Hawking noticed that no more than 29\% of the initial BH mass could be extracted via superradiance~\cite{Hawking:1971tu}. Yet, the simulations by East and Pretorius, which followed the growth of the dominant superradiant mode, only reached about $9\%$. Was this because of the choice of parameters? Could the value be much closer to 29\%? Is there a difference between scalar and vector superradiance concerning this maximal efficiency?

In this letter, we shall provide a simple argument that for both the scalar and vector case there is a roughly similar bound of around $10\%$ for the maximal energy extraction due to the dominant superradiant mode from a Kerr BH; thus the efficiency is $\epsilon\lesssim 0.1$,\footnote{$\epsilon$ is the fraction of the initial mass that is extracted in the process.} regardless the spin of the bosonic field. This bound is based on: $(i)$ a scanning of the BHs solutions with synchronised scalar and Proca hair that can result from the growth of the dominant superradiant mode; $(ii)$ the rationale that this superradiant evolution is approximately conservative, which is supported by the evolutions in~\cite{East:2017ovw}. Under this assumption, superradiance merely redistributes the mass and angular momentum of a Kerr BH, splitting it amongst the trapped region  and the bosonic hair. Since Kerr BHs have a dimensionless spin $j$ that obeys $j\leqslant 1$ (Kerr bound), the corresponding hairy BH that forms after the superradiant growth from Kerr must also obey this bound. We observe that generic hairy BHs \textit{do not} obey $j\leqslant 1$~\cite{Herdeiro:2014goa}. 
Scanning the parameter space, we can identify the hairiest solutions, under the constraint $0\leqslant j\leqslant 1$. These occur precisely for hairy BHs with $j=1$ and at a certain $M\mu$, which (slightly) depends on the spin of the field. Identifying the fraction of energy in the bosonic field in these solutions with the maximal efficiency, we obtain:
\begin{equation}
{\bf Scalar: } \ \ \epsilon_{\rm max}\sim 0.099 \  \ ({\rm at} \ M\mu\approx0.24) \  ; \qquad 
{\bf Vector: } \ \ \epsilon_{\rm max}\sim 0.104 \ ({\rm at} \ M\mu\approx0.27)  \ .
\label{result}
\end{equation}
In particular we observe that, for the vector case, the maximal energy extraction reported in~\cite{East:2017ovw}, of $\sim 9\%$, occured for  $M\mu=0.25$, which shows an interesting agreement with~\eqref{result}. Equation~\eqref{result} is the main result in this letter. It \textit{predicts} that numerical evolutions (yet to be performed) similar to those in~\cite{East:2017ovw} for the scalar case will lead to a similar result for the maximal efficiency, smaller than $10\%$.  In the remaining of this paper we shall detail how the result~\eqref{result} is obtained.

\section{\label{sec:level2}Black holes with synchronised hair}

BHsSH are families of four--dimensional, asymptotically--flat, stationary solutions of Einstein’s gravity minimally coupled to a complex bosonic field $\psi$ with non--vanishing mass $\mu$. The bosonic field can be a scalar (first discussed in~\cite{Herdeiro:2014goa}) or a vector (first discussed in~\cite{Herdeiro:2016tmi}). Such spacetimes are regular on and outside an event horizon. The simplest solutions arise for free bosonic fields, but generalization with interacting fields and/or non--minimal couplings are possible and have been studied, $e.g.$~\cite{Herdeiro:2015tia}.
Besides the field's mass, which defines a scale and is set in the action, the space of solutions is conveniently characterised as a two--dimensional domain,  spanned by two continuous dimensionless parameters: the ADM mass in units of the field's mass\footnote{For clarity we remark that, reinserting units, this dimensionless parameter is $M\mu/M_{\rm Pl}^2$, where $M_{\rm Pl}$ is the Planck mass.}, $M\mu$, and the oscillation frequency of the matter field in units of the field's mass, $\omega/\mu$. For each value of $(\omega/\mu, M\mu)$ there is a single BH solution in a certain two--dimensional domain -- see~\autoref{fig:1}. The range of \textit{physical} masses $M$ (say in solar masses) becomes defined after specifying the scale $\mu$.

Actually, the continuous family of solutions just described is only one amongst an infinite discrete set of such continuous families of hairy BHs. This discrete set is labeled by two integers: the number of nodes of the appropriate radial functions, $n\in \mathbb{N}_0$, and the azimuthal harmonic index, $m\in\mathbb{Z}^+$, which, like $\omega$, enters the bosonic field ansatz, $\psi\sim e^{-i\omega t+i m \varphi}$. Both $n$ and $m$ can be seen as excitation numbers. Here we shall focus on the fundamental solutions with $(n,m)=(0,1)$, which are the ones that naturally emerge as the equilibrium configurations from the growth of the dominant superradiant mode in Kerr~\cite{Herdeiro:2017phl}, but some of the excited solutions have also been explicitly constructed -- see $e.g.$~\cite{Wang:2018xhw,Delgado:2019prc}. 

BHsSH rely on a synchronisation between the event horizon's angular velocity $\Omega_H$ and the field's phase angular velocity $\omega/m$, \textit{i.e.} they satisfy the \textit{synchronisation condition} 
\begin{align}
\Omega_H=\frac{\omega}{m}\ .
\label{eq2}
\end{align}
Thus, the $(\omega/\mu, M\mu)$ parameterization of the domain of existence can be equally seen as a $(m\Omega_H/\mu, M\mu)$ parameterization, which is a set of more physically intuitive quantities. Each solution in this domain has two extra global quantities, besides the ADM mass: the total angular momentum, $J\mu^2$, and the Noether charge associated with the global $U(1)$ symmetry provided by the  complex nature of the bosonic field, $Q\mu
^2$. Unlike the mass and angular momentum, the Noether charge is not associated with a Gauss law, meaning it cannot be measured by an observer at infinity.
Since the domain of existence is two--dimensional, the three global quantities $(M\mu,J\mu^2,Q\mu^2)$ are not independent, but no simple relation between them is known.

 The global charges $M$ and $J$ can be expressed as $M=M_H+M_{\psi}$ and $J=J_H+J_\psi$, where $M_H$ and $J_H$ ($M_\psi$ and $J_\psi$) are the energy and angular momentum inside (outside) the event horizon, respectively. These are rigorously defined by Komar integrals -- see $e.g.$~\cite{Herdeiro:2015gia,Herdeiro:2016tmi}. It is convenient to define the dimensionless total and horizon angular momenta, $j\equiv J/M^2$ and $j_H\equiv J_H/M_H^2$, respectively. BHsSH can violate the Kerr bound, in terms of asymptotic and/or horizon quantities \cite{Delgado:2016zxv}, and in fact do so in large extensions of their domain of existence, although their horizon \textit{linear} velocity never exceeds the speed of light~\cite{Herdeiro:2015moa}.

The proportion of energy and angular momentum in the bosonic field (\textit{i.e.} outside the event horizon), for a given solution, is measured by the fractions
\begin{align}
p\equiv \frac{M_\psi}{M}\ ,
\quad
q\equiv \frac{J_\psi}{J}\ .
\label{eq1}
\end{align}
The quantities $p$ and $q$ measure the \textit{hairiness} of the solutions. Note that $p,q\in [0,1]$. 
They reduce to Kerr BHs in equilibrium with linearised bosonic fields when $p,q\to0$ (\textit{Kerr limit}) and to spinning bosonic stars when $p,q\to1$ (\textit{solitonic limit}).  If a hairy BH is the equilibrium state obtained from the superradiance instability of Kerr, and under the aforementioned assumption of an approximately conservative process, then we identify the efficiency of the process as $\epsilon=p$.

\section{\label{sec:level2.2}Domain of existence}

\begin{figure}[h]
\centering
\includegraphics[width=0.475\columnwidth]{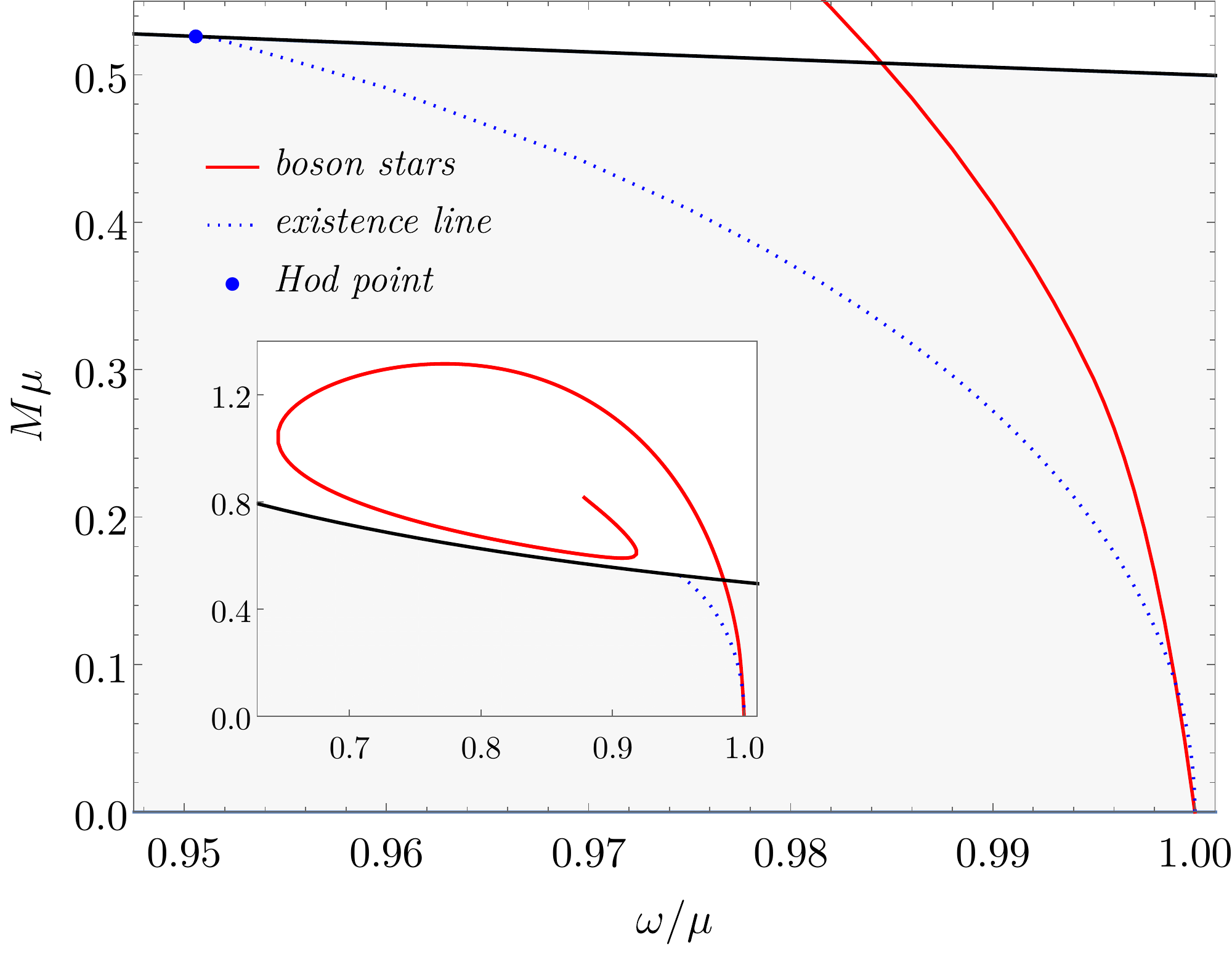}
\quad
\includegraphics[width=0.475\columnwidth]{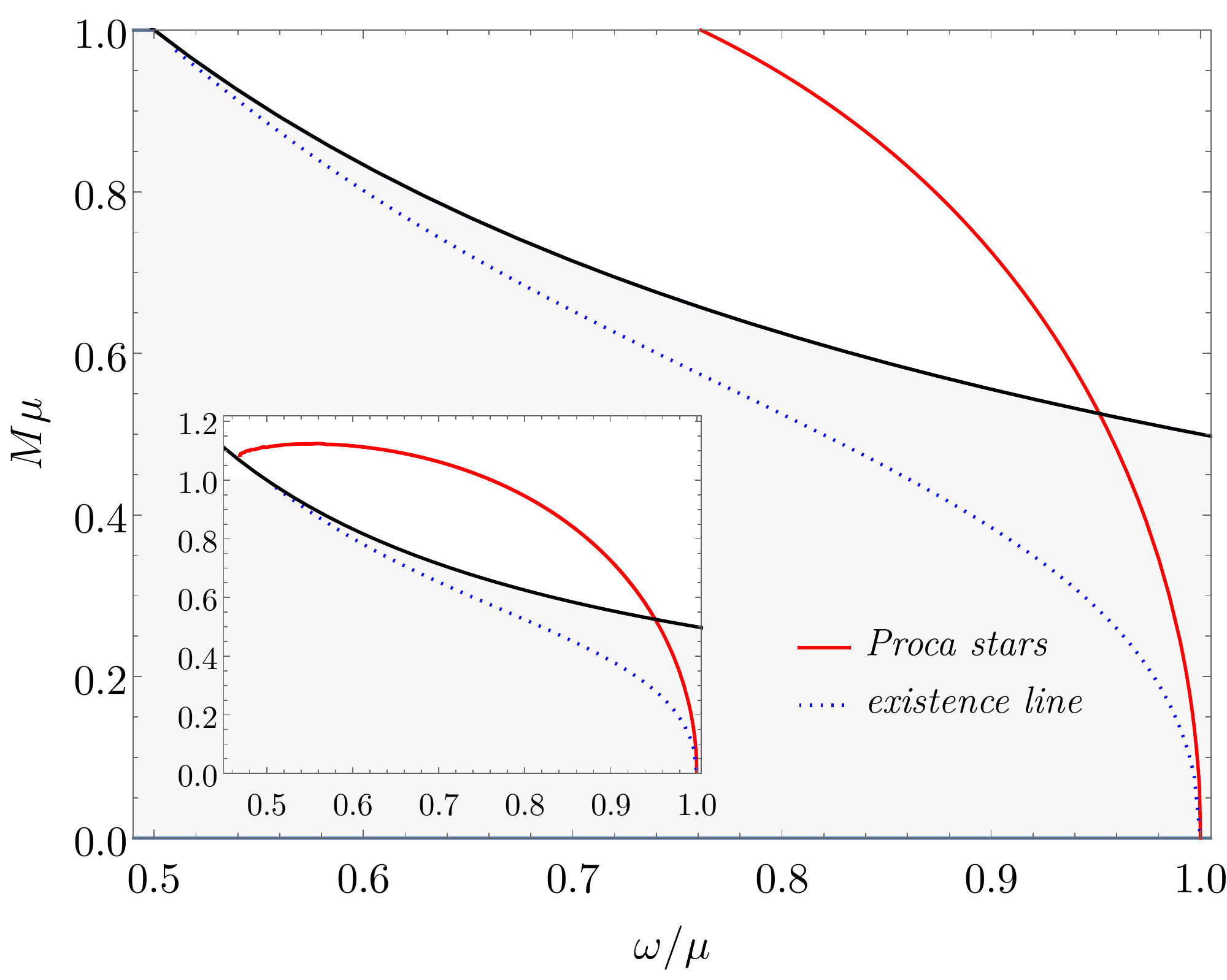}
\caption{Region of interest (cf. \autoref{sec:level2.4}) of the domain of existence of BHs with synchronised scalar (left panel) and vector (right panel) hair with $(n,m)=(0,1)$ in the $M\mu$ \textit{vs.} $\omega/\mu$ plane. The insets in both panels show the full domain of existence. }
\label{fig:1}
\end{figure}

\autoref{fig:1} shows the domain of existence of Kerr BHs with synchronised scalar (left panel) and vector (right panel) hair with $(n,m)=(0,1)$.
A detailed comparison between the two families can be found 
in \cite{Santos:2020pmh}. The light gray shaded region represents the domain of existence of Kerr BHs in Einstein's gravity, which satisfy the Kerr bound, \textit{i.e.} $j\leqslant 1$. Solutions saturating this bound fall into the black solid line. The domain of existence of BHsSH is bounded by: $(i)$ the \textit{existence line} (blue dotted line), a line segment comprised of solutions describing bound states between Kerr BHs and linearised bosonic fields ($p=q=0$). This line segment joins the \textit{Minkowski limit} ($M,J\to0$) to the Kerr bound line. It is half--open, including the upper endpoint only. The latter is known as the \textit{Hod point} \cite{Hod:2012px} and can be found analytically for the scalar case (blue point in the left panel of \autoref{fig:1}). 
And $(ii)$ the \textit{bosonic star line} (red solid line), comprised of solutions describing spinning bosonic stars ($p=q=1$). 

As already discussed, the (dominant mode) superradiant instability of Kerr may form some of the hairy BHs in this domain of existence. 
 This phenomenon occurs whenever the Kerr solution is exposed to a scalar field perturbation for which the field's phase angular velocity satisfies the \textit{superradiant condition} $\omega/m<\Omega_H$. For a massless field, the rotational energy is radiated to infinity, leaving a Kerr BH with lower mass and angular momentum, in fact decreasing $j$. But for a bosonic field with non--vanishing mass and with a Compton wavelength comparable to, or larger than, the Kerr BH, a \textit{superradiant instability} sets in, driving the configuration to a new equilibrium state (for complex fields). 

Kerr BHs which lie to the right of the  existence line are unstable when they are perturbed by linearized bosonic fields with the $m=1$ azimuthal mode. The seed solutions of the evolutions described in \cite{East:2017ovw} were Kerr BHs with initial mass $M$ and dimensionless angular momentum $j=j_H=0.99$ in the presence of a vector field with mass $M\mu\in\{0.25,0.30,0.40,0.50\}$ and azimuthal number $m=1$. These Kerr BHs  gradually developed hair, attaining equilibrium when the horizon and the field synchronised. The corresponding migrations can be observed in Fig.~5 in~\cite{Herdeiro:2017phl}.

\newpage

\section{\label{sec:level2.3} Analytic bounds on ``hairiness" for $j\leqslant 1$}

Hawking's area theorem sets an upper limit of $29\%$ for the efficiency of energy extraction from Kerr BHs by superradiance. This is simple to see. The horizon area of a Kerr BH with mass $M_H$ and angular momentum $J_H$ is $A=8\pi M_H^2(1+\sqrt{1-j_H^2})$
(we recall that
$M=M_H$,
$J=J_H$
for a vacuum Kerr BH). The irreducible mass $M_\text{irr}$ is defined as the mass of the Schwarzschild BH that results when all angular momentum has been extracted by reversible transformations, \textit{i.e.} leaving the horizon area unchanged. The area theorem ($\delta A\geqslant0$) then dictates that $\delta M_\text{irr}\geqslant0$. Like the area of the horizon, the irreducible mass remains unchanged (increases) in (ir)reversible transformations. 

Starting from a Kerr BH, the maximum amount of energy that can be extracted from it is
\begin{align}
M_H-M_\text{irr}=M_H\left[1-\frac{1}{\sqrt{2}}\left(1+\sqrt{1-j_H^2}\right)^{1/2}\right]\ ,
\end{align}
resulting in a Schwarzschild BH with mass $M_\text{irr}$. It is therefore possible to extract up to $1-1/\sqrt{2}\approx 29\%$ of the energy, with the upper limit corresponding to an initial extremal Kerr BH ($j_H=1$) and a final Schwarzschild BH. Of thermodynamic nature, this limit applies to any (reversible or irreversible) transformation whereby rotational energy is extracted from a Kerr BH, including superradiance. 

In light of this maximal theoretical efficiency, $\epsilon \lesssim 0.29$, BHsSH grown from superradiance have $p\lesssim 0.29$. 
 The simulations in \cite{East:2017ovw} showed, however, only up to $9\%$ of the initial energy is transferred into the (vector) field. Furthermore, they exhibit negligible dissipation. This suggests that the evolution of superradiant instabilities is nearly conservative, \textit{i.e.} preserves the energy $M$ and angular momentum $J$, thus leaving $j$ almost unchanged. Accordingly, $j\leqslant1$ should be satisfied throughout the evolution, since it is satisfield by the initial (Kerr BH) state.

\begin{figure}[h]
\centering
\includegraphics[width=0.475\columnwidth]{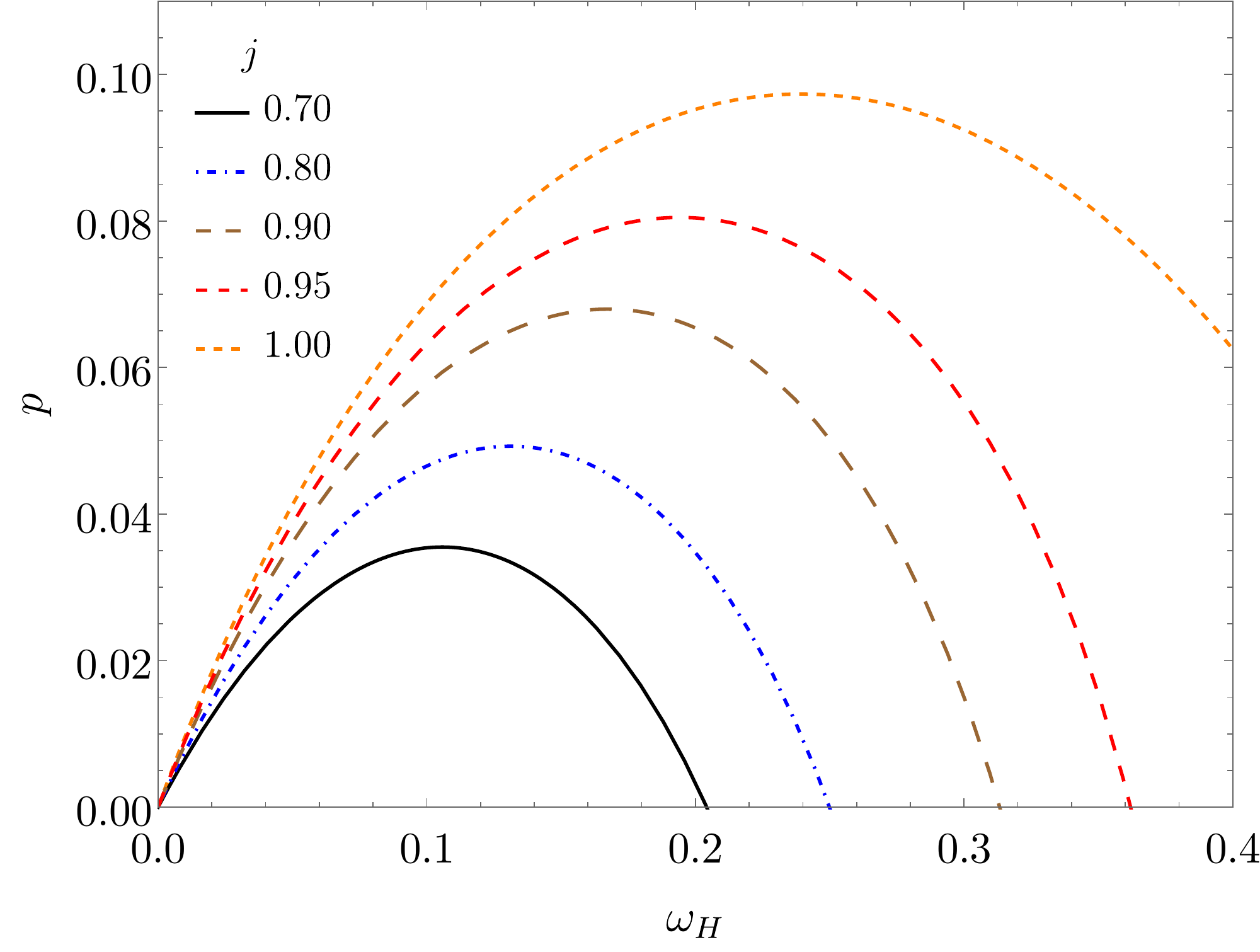}
\caption{Fraction of energy in the bosonic hair $p$ according to relation~\eqref{analytic}.}
\label{fig:0}
\end{figure}

 It was already suggested in \cite{Herdeiro:2017phl} that this upper limit on $j$ places tighter constraints on the hairiness than the thermodynamic limit.  This was done using an analytical model proposed therein to describe physical quantities of the hairy BHs which are Kerr--like. According to this model,  BHsSH which are sufficiently Kerr--like have a fraction of energy in the bosonic field $p$ obeying 
\begin{align}
p=1+\frac{1-\sqrt{1-16\omega_H^2(j\omega_H-1)^2}}{8\omega_H^2(j\omega_H-1)}\ ,
~~{\rm where}~~\omega_H=M \Omega_H\ .
\label{analytic}
\end{align}
In Fig.~\ref{fig:0} we show how $p$ derived from this relation varies with $j$ and $\omega_H$.
The analysis shows that $j\leqslant 1$ implies $p\lesssim 0.10$,  approximately one third of the thermodynamic bound. 

\section{\label{sec:level2.4} A bound on ``hairiness" from scanning BHsSH}

\begin{figure}[h]
\centering
\includegraphics[width=0.475\columnwidth]{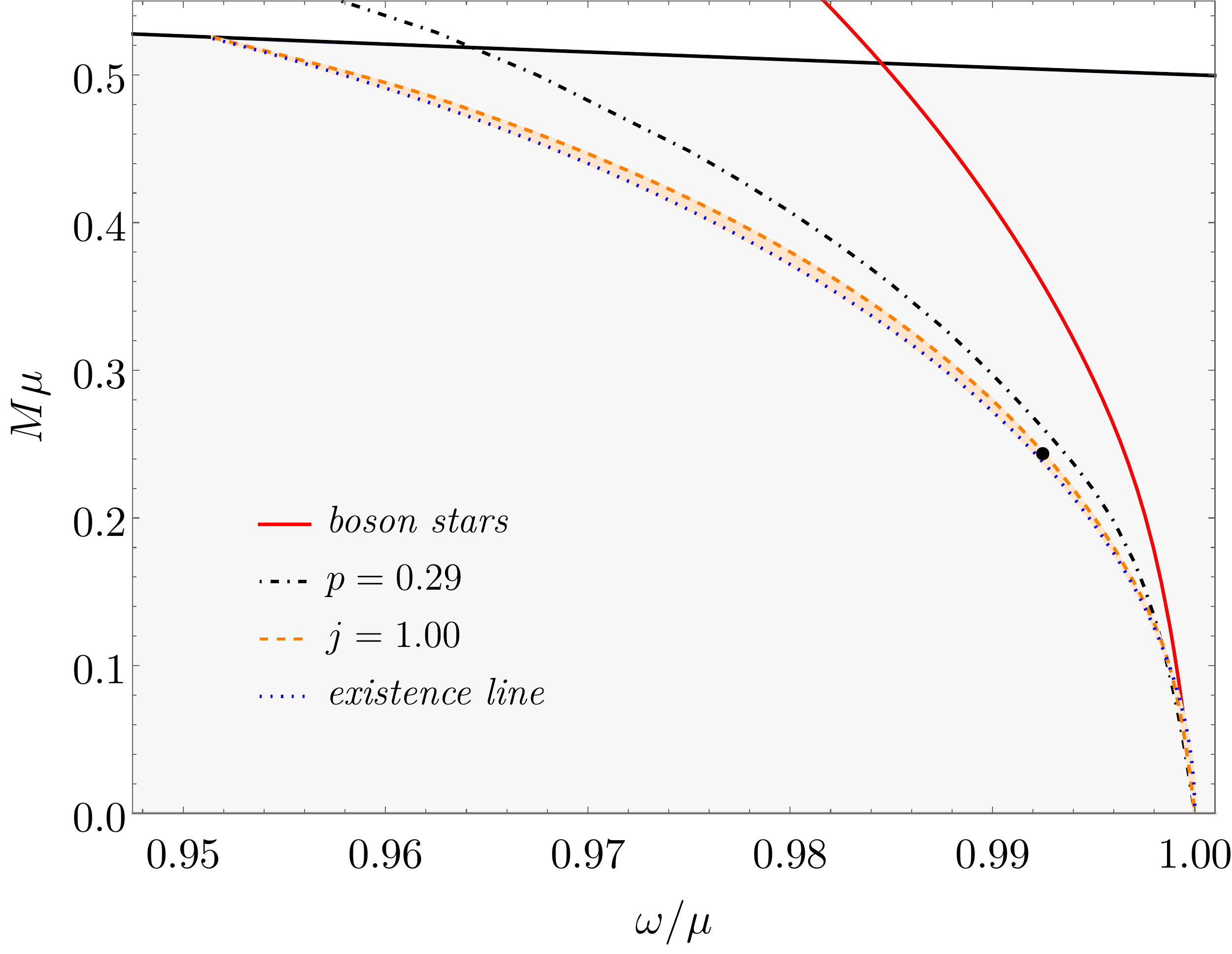}
\quad
\includegraphics[width=0.475\columnwidth]{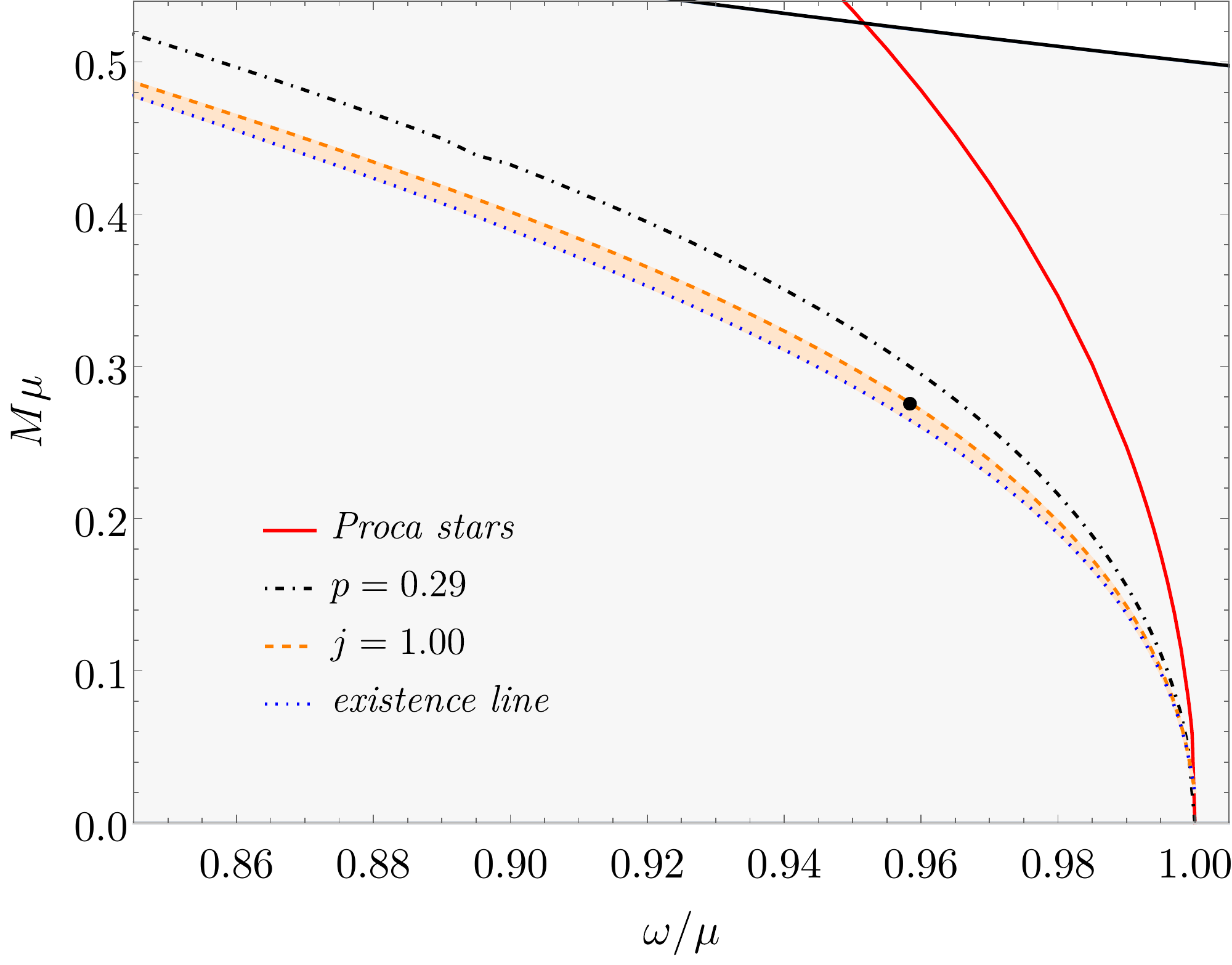}
\caption{Same as in \autoref{fig:1}, but with two additional lines: the dash--dotted black line separates BHsSH with
less (to the left) and more (to the right) than $29\%$ of the total energy in bosonic field, whereas the dashed orange line separates BHsSH satisfying (to the left) and violating (to the right) the Kerr bound. The light orange shaded region comprises BHsSH which satisfy this bound. The black circles represent the ``hairiest" solutions in the region of interest (cf. \autoref{tab:2}).}
\label{fig:2}
\end{figure}

Instead of using an approximate analytical model, we can in fact use the data on hairy BHs to see how large is $p$ for $j\leqslant 1$. The region of interest, containing BHsSH that could emerge from the superradiant instability of Kerr BHs, is a subset of the domain of existence (shaded light orange in \autoref{fig:2}), bounded by two lines: the existence line and the $j=1$ line. These lines meet at the Hod point, which corresponds to an extremal Kerr BH ($j=j_H=1$). The ``hairiness" trend is that, for fixed $M\mu$, $p$ increases as $\omega/\mu$ increases. Since the $p=0.29$ line always lies to the right of the $j=1$ line, the latter sets a tighter (frequency--dependent) upper limit on the hairiness than the former, as expected.

 \autoref{tab:1} lists the properties of BHsSH with $j=1$ and $M\mu\in\{0.25,0.30,0.40,0.50\}$ for both the scalar and vector cases. These illustrative values were chosen to match the ones taken in the simulations in~\cite{East:2017ovw}, which were carried out for the vector case. According to our assumption that $\epsilon=p$, these are the hairiest solutions with such masses that can be formed from superradiance. For instance, a Kerr BH with synchronised scalar (vector) hair has at most about $9.7\%$ ($10.4\%$) of its energy in the field when $M\mu=0.25$. The results in \autoref{tab:1} are compatible with the findings reported in \cite{East:2017ovw}, suggesting in particular that the superradiant amplification of the (vector) field in the maximal efficiency case reported therein, $M\mu=0.25$, is (approximately) as efficient as it can be.

\begin{table}[h!]
\centering
\begin{tabular}{c||c||cccc}
& $M\mu$ & $\omega/\mu$ & $M\omega$ & $p$ & $q$\\ \midrule
\multirow{4}{*}{\rotatebox[origin=c]{90}{\textbf{Scalar}}} & $0.25$ & 0.9921 & 0.2480 & 0.0971 & 0.3856\\
& $0.30$ & 0.9884 & 0.2965 & 0.0951 & 0.3132\\
& $0.40$ & 0.9774 & 0.3910 & 0.0686 & 0.1669\\
& {$0.50$} & 0.9587 & 0.4794 & 0.0160 & 0.0303\\ \midrule
\multirow{4}{*}{\rotatebox[origin=c]{90}{\textbf{Vector}}} & $0.25$ & 0.9667 & 0.2417 & 0.1035 & 0.3984\\
& $0.30$ & 0.9496 & 0.2849 & 0.1038 & 0.3259\\
& $0.40$ & 0.9009 & 0.3604 & 0.0933 & 0.2057\\
& $0.50$ & 0.8356 & 0.4178 & 0.0738 & 0.1221\\
\end{tabular}
\caption{``Hairiness" of BHsSH with $(n,m)=(0,1)$ and $j=1$, for selected values of $M\mu$.
}
\label{tab:1}
\end{table}

\begin{figure}[h]
\centering
\includegraphics[width=0.475\columnwidth]{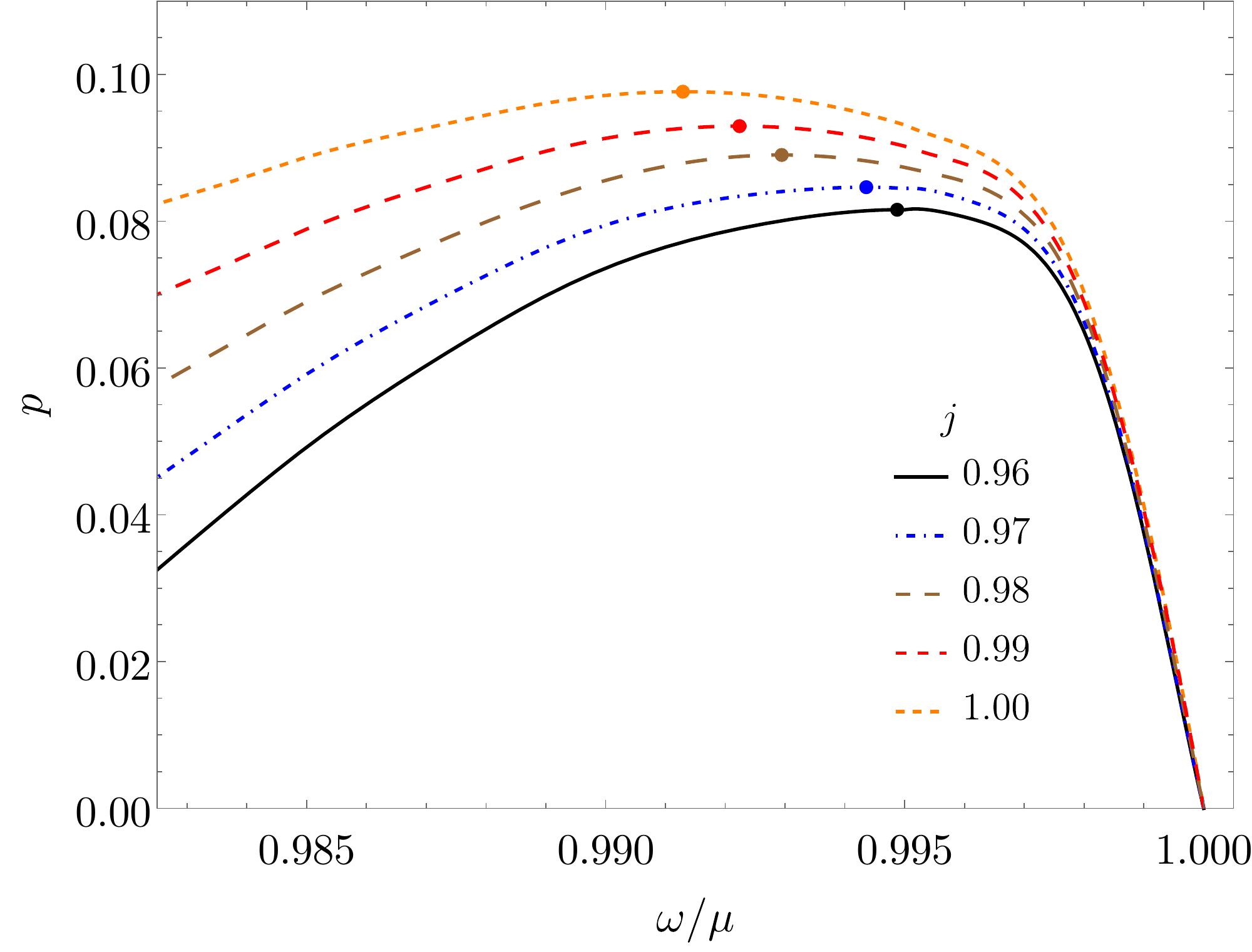}
\quad
\includegraphics[width=0.475\columnwidth]{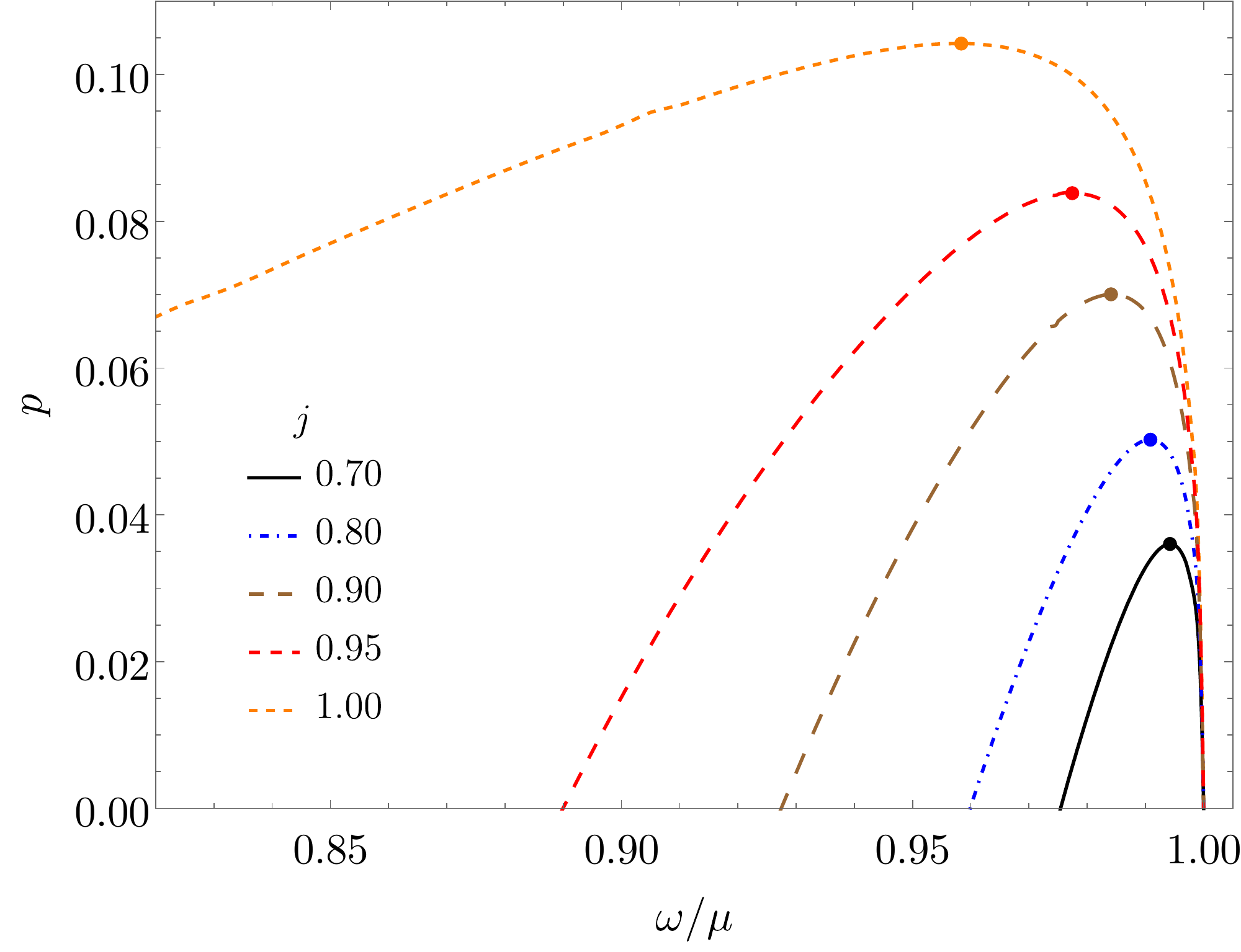}\\
\includegraphics[width=0.475\columnwidth]{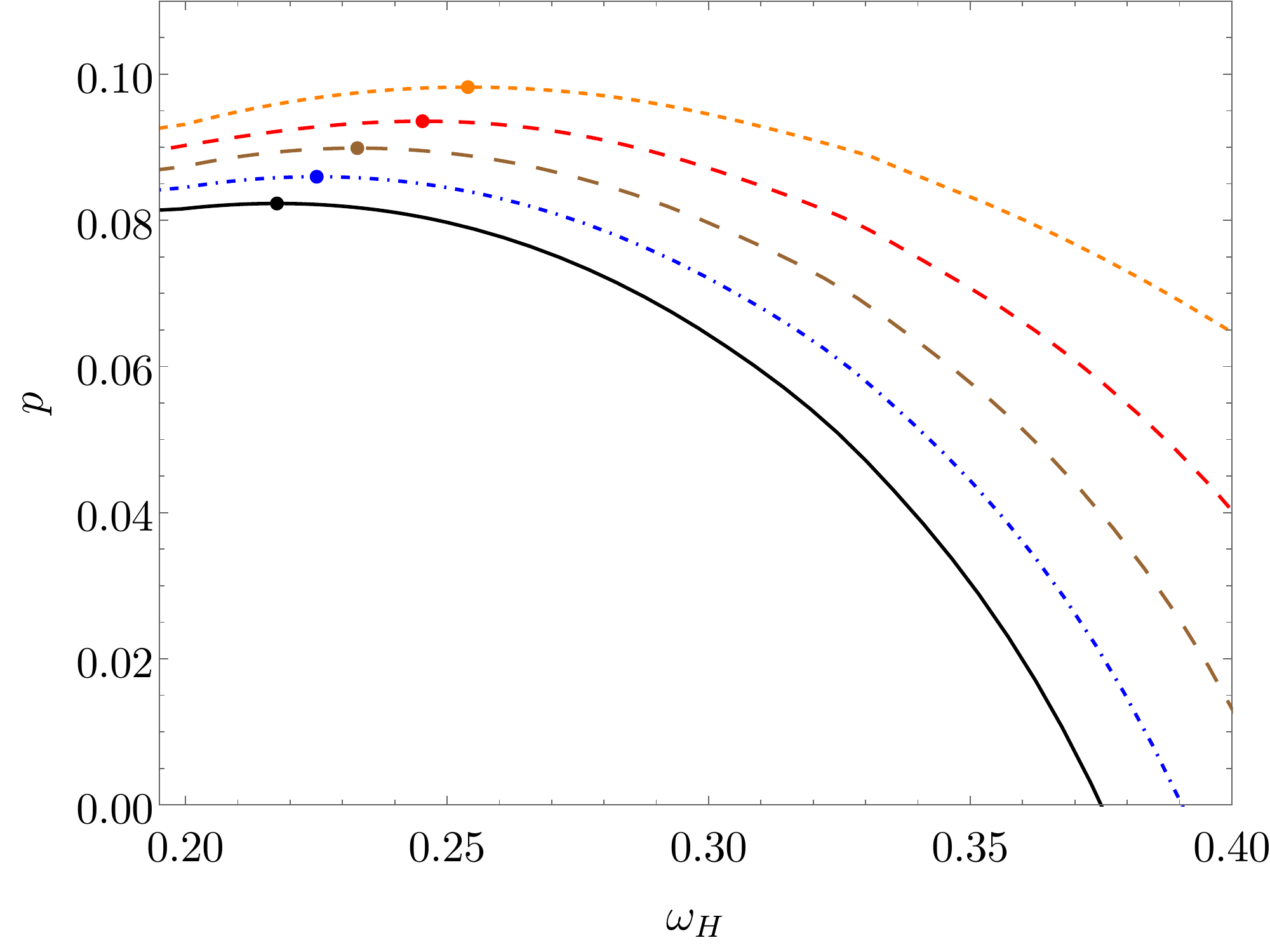}
\quad
\includegraphics[width=0.475\columnwidth]{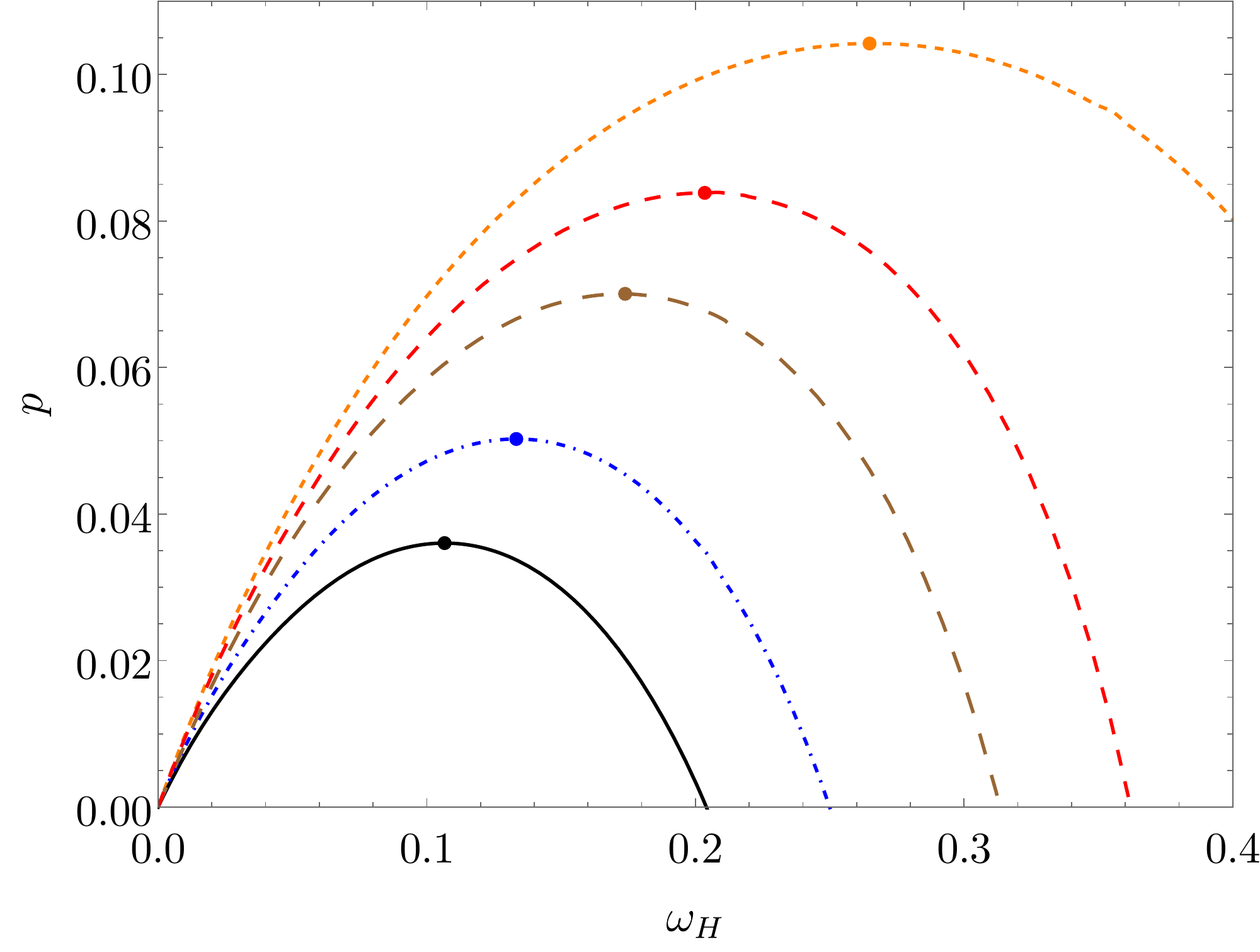}
\caption{Fraction $p$ of the total energy contained in the bosonic field of BHs with synchronised scalar (left panels) and vector (right panels) hair for selected values of $j$. The top (bottom) panels show the dependence of $p$ on $\omega/\mu$ ($\omega_H=M\Omega_H$). The circles pinpoint the corresponding maximum.}
\label{fig:3}
\end{figure}

A more comprehensive analysis is provided in  \autoref{fig:3}.
Starting from the Hod point, $p$ increases as one moves downstream along the $j=1$ line, reaching a maximum and then decreasing towards the Minkowski limit. Solutions with fixed $j$ values below unity show a similar behavior. The maximum occurs at larger (lower) values of $\omega/\mu$ ($\omega_H$) as $j$ decreases. The global maximum of $p$ occurs at $M\mu\approx0.24$ ($0.27$) and it is about $0.099$ ($0.104$) in the scalar (vector) case, as shown in \autoref{tab:2}, corresponding to the values reported in Eq.~\eqref{result}. This suggests the maximal efficiency is not very sensitive to the spin of the bosonic field.  The corresponding solutions satisfy the Kerr bound in terms of horizon quantities as well. \autoref{tab:2} also shows they are well described by the analytical model introduced in \cite{Herdeiro:2017phl}, valid for any bosonic field. 

\begin{table}[h!]
\centering
\begin{tabular}{c||cccccc}
\toprule
\textbf{Model} & $M\mu$ & $\omega/\mu$ & $M\omega$ & $p$ & $q$ & $j_H$\\ \midrule
Scalar \cite{Herdeiro:2014goa} & 0.2445 & 0.9925 & 0.2426 & 0.0989 & 0.4010 & 0.7367\\
Vector \cite{Herdeiro:2016tmi} & 0.2761 & 0.9584 & 0.2646 & 0.1042 & 0.3621 & 0.7981\\
Analytical \cite{Herdeiro:2017phl} & -- & -- & 0.2393 & 0.0973 & 0.4067 & 0.7282\\
\bottomrule
\end{tabular}
\caption{Properties of the ``hairiest" BHsSH with $(n,m)=(0,1)$ which are comparable to Kerr BHs ($i.e.$ obey $j\leqslant 1$).  These are characterised by $j=1$ and are pinpointed as black circles in \autoref{fig:2}.}
\label{tab:2}
\end{table}

\section{\label{sec:level3}Remarks}

Some final remarks are in order. 
Firstly, this new (tighter) upper bound on the ``hairiness" of BHsSH grown from superradiance (those represented by the region of interest) does not exclude the possibility that hairier BHs can appear from other dynamical formation channels, such as the merger of bosonic stars \cite{Sanchis-Gual:2020mzb}. That is, the region of \textit{dynamically viable solutions}, corresponding to those that can form by some mechanism and be sufficiently long--lived, can go beyond the region of interest discussed here.

Secondly, BHsSH in the region of interest can be arbitrarily close to Kerr BHs and therefore are quite Kerr--like: they are more accurately described as event horizons surrounded by a bosonic cloud rather than bosonic stars with an event horizon at its center \cite{Santos:2020pmh}. For instance, the areal radius of their shadow is at most $11\%$ larger than that of comparable Kerr BHs \cite{Cunha:2019ikd}. 

Thirdly, like Kerr BHs, BHsSHs are prone to their own superradiant instability \cite{Ganchev:2017uuo}. At fixed $m$, they are unstable against bosonic field modes with $\tilde{m}>m$. For constant $M\mu$, the strength of the instability decreases as the BHs become hairier (\textit{i.e.} as one moves away from the existence line) \cite{Degollado:2018ypf}. In the region of interest, it is minimum for BHs on the $j=1$ line. If the timescale of the instability is larger than the age of the Universe, the hairy BH is effectively stable \cite{Degollado:2018ypf}. Effective stability is expected to occur for $M\mu\lesssim0.25$. Interestingly, the ``hairiest" BHs are characterised by $M\mu\approx 0.25$ (see \autoref{tab:2}) and thus might be stable on cosmological timescales, for the appropriate mass range.  

Finally, let us comment on three potential limitations of our approach. 
The first one is that we have assumed that the evolution from Kerr into a hairy BH is conservative, unaltering $j$. If that would not be the case, and some of the system's energy is dissipated towards infinity by gravitational waves or gravitational cooling~\cite{Seidel:1993zk} ($i.e.$ ejection of the bosonic field), would this challenge the bound? For the Kerr BH, superradiance with dissipation ($e.g.$ via the scattering of a massless bosonic field), has a net effect of increasing the reduced area $A/M^2$. This implies a reduction of $j$. Thus, it seems likely that also for the hairy BHs any dissipation will reduce $j$ further. Accordingly, the bound obtained under the assumption of a non--dissipative evolution is a robust, conservative bound. 
The second potential limitation is that we have obtained the bound~\eqref{result} from a specific set of solutions of BHsSH, namely those of the simplest bosonic model, without self--interactions. If we allow for self--interactions, is the bound significantly affected? Since the bosonic field is small in the region of interest, as it is very close to the existence line (cf.\ \autoref{fig:2}), any non--linearities will be negligible. Thus, we expect this bound to be universal, in the sense of also applying to models with generic self--interactions. 
 Indeed,  a preliminary investigation of the results reported in \cite{Herdeiro:2015tia} confirms this expectation.
That work studied BHsSH in a model with a massive complex scalar field with a $quartic$ self--interaction, while
the case of synchronised BH solutions with a self--interacting vector field has not yet been considered in the literature.
 As a final possible limitation, we have only considered isolated BHs, avoiding the issue of accretion. Accretion is known to spin up BHs and thus it may counter--act the effect of superradiance -- see $e.g.$~\cite{Brito:2014wla,Sanchis-Gual:2020mzb}. Our bound, however, is obtained for BHs with $j=1$. For the Kerr case, this means no further accretion is possible. A similar sharp statement cannot be applied for the hairy BHs, since $j$ can exceed unity. Yet, since these are Kerr--like BHs, it seems plausible that the impact of accretion may be small. 

\section*{Acknowledgements}

This work is supported  by the Center for Astrophysics and Gravitation (CENTRA) and by the Center for Research and Development in Mathematics and Applications (CIDMA) through the Portuguese Foundation for Science and Technology (FCT -- Funda\c{c}\~ao para a Ci\^encia e a Tecnologia), references UIDB/00099/2020, UIDB/04106/2020 and UIDP/04106/2020,
 and by national funds (OE), through FCT, I.P., 
in the scope of the framework contract foreseen in the numbers 4, 5 and 6 of the article 23,of the Decree-Law 57/2016, of August 29, changed by Law 57/2017, of July 19.  
The authors acknowledge support  from the projects PTDC/FIS-OUT/28407/2017, CERN/FIS-PAR/0027/2019 and PTDC/FIS-AST/3041/2020. This work has further been supported by  the  European  Union's  Horizon  2020  research  and  innovation  (RISE) programme H2020-MSCA-RISE-2017 Grant No.~FunFiCO-777740. The authors would like to acknowledge networking support by the COST Action CA16104.

  \bibliographystyle{ieeetr}  
  \bibliography{references}

\begin{thebibliography}{10}

\bibitem{Akiyama:2019cqa}
K.~Akiyama {\em et~al.}, ``{First M87 Event Horizon Telescope Results. I. The
  Shadow of the Supermassive Black Hole},'' {\em Astrophys. J.}, vol.~875,
  no.~1, p.~L1, 2019.

\bibitem{Abbott:2016blz}
B.~P. Abbott {\em et~al.}, ``{Observation of Gravitational Waves from a Binary
  Black Hole Merger},'' {\em Phys. Rev. Lett.}, vol.~116, no.~6, p.~061102,
  2016.

\bibitem{LIGOScientific:2018mvr}
B.~P. Abbott {\em et~al.}, ``{GWTC-1: A Gravitational-Wave Transient Catalog of
  Compact Binary Mergers Observed by LIGO and Virgo during the First and Second
  Observing Runs},'' {\em Phys. Rev. X}, vol.~9, no.~3, p.~031040, 2019.

\bibitem{LIGOScientific:2020ibl}
R.~Abbott {\em et~al.}, ``{GWTC-2: Compact Binary Coalescences Observed by LIGO
  and Virgo During the First Half of the Third Observing Run},'' {\em Phys.
  Rev. X}, vol.~11, p.~021053, 2021.

\bibitem{Kerr:1963ud}
R.~P. Kerr, ``{Gravitational field of a spinning mass as an example of
  algebraically special metrics},'' {\em Phys. Rev. Lett.}, vol.~11,
  pp.~237--238, 1963.

\bibitem{Carter:1971zc}
B.~Carter, ``{Axisymmetric Black Hole Has Only Two Degrees of Freedom},'' {\em
  Phys. Rev. Lett.}, vol.~26, pp.~331--333, 1971.

\bibitem{Robinson:1975bv}
D.~C. Robinson, ``{Uniqueness of the Kerr black hole},'' {\em Phys. Rev.
  Lett.}, vol.~34, pp.~905--906, 1975.

\bibitem{Chrusciel:2012jk}
P.~T. Chrusciel, J.~Lopes~Costa, and M.~Heusler, ``{Stationary Black Holes:
  Uniqueness and Beyond},'' {\em Living Rev. Rel.}, vol.~15, p.~7, 2012.

\bibitem{Bekenstein:1972ny}
J.~D. Bekenstein, ``{Transcendence of the law of baryon-number conservation in
  black hole physics},'' {\em Phys. Rev. Lett.}, vol.~28, pp.~452--455, 1972.

\bibitem{Herdeiro:2015waa}
C.~A.~R. Herdeiro and E.~Radu, ``{Asymptotically flat black holes with scalar
  hair: a review},'' {\em Int. J. Mod. Phys. D}, vol.~24, no.~09, p.~1542014,
  2015.

\bibitem{Ruffini:1971bza}
R.~Ruffini and J.~A. Wheeler, ``{Introducing the black hole},'' {\em Phys.
  Today}, vol.~24, no.~1, p.~30, 1971.

\bibitem{Penrose:1964wq}
R.~Penrose, ``{Gravitational collapse and space-time singularities},'' {\em
  Phys. Rev. Lett.}, vol.~14, pp.~57--59, 1965.

\bibitem{Hawking:1970zqf}
S.~W. Hawking and R.~Penrose, ``{The Singularities of gravitational collapse
  and cosmology},'' {\em Proc. Roy. Soc. Lond. A}, vol.~314, pp.~529--548,
  1970.

\bibitem{Brito:2015oca}
R.~Brito, V.~Cardoso, and P.~Pani, ``{Superradiance}: {New Frontiers in Black
  Hole Physics},'' {\em Lect. Notes Phys.}, vol.~906, pp.~pp.1--237, 2015.

\bibitem{Arvanitaki:2010sy}
A.~Arvanitaki and S.~Dubovsky, ``{Exploring the String Axiverse with Precision
  Black Hole Physics},'' {\em Phys. Rev. D}, vol.~83, p.~044026, 2011.

\bibitem{Ng:2020ruv}
K.~K.~Y. Ng, S.~Vitale, O.~A. Hannuksela, and T.~G.~F. Li, ``{Constraints on
  Ultralight Scalar Bosons within Black Hole Spin Measurements from the
  LIGO-Virgo GWTC-2},'' {\em Phys. Rev. Lett.}, vol.~126, no.~15, p.~151102,
  2021.

\bibitem{Yuan:2021ebu}
C.~Yuan, R.~Brito, and V.~Cardoso, ``{Probing ultralight dark matter with
  future ground-based gravitational-wave detectors},'' {\em Phys. Rev. D},
  vol.~104, no.~4, p.~044011, 2021.

\bibitem{East:2017ovw}
W.~E. East and F.~Pretorius, ``{Superradiant Instability and Backreaction of
  Massive Vector Fields around Kerr Black Holes},'' {\em Phys. Rev. Lett.},
  vol.~119, no.~4, p.~041101, 2017.

\bibitem{Herdeiro:2017phl}
C.~A.~R. Herdeiro and E.~Radu, ``{Dynamical Formation of Kerr Black Holes with
  Synchronized Hair: An Analytic Model},'' {\em Phys. Rev. Lett.}, vol.~119,
  no.~26, p.~261101, 2017.

\bibitem{Herdeiro:2016tmi}
C.~Herdeiro, E.~Radu, and H.~R\'unarsson, ``{Kerr black holes with Proca
  hair},'' {\em Class. Quant. Grav.}, vol.~33, no.~15, p.~154001, 2016.

\bibitem{Herdeiro:2014goa}
C.~A.~R. Herdeiro and E.~Radu, ``{Kerr black holes with scalar hair},'' {\em
  Phys. Rev. Lett.}, vol.~112, p.~221101, 2014.

\bibitem{Ganchev:2017uuo}
B.~Ganchev and J.~E. Santos, ``{Scalar Hairy Black Holes in Four Dimensions are
  Unstable},'' {\em Phys. Rev. Lett.}, vol.~120, no.~17, p.~171101, 2018.

\bibitem{Degollado:2018ypf}
J.~C. Degollado, C.~A.~R. Herdeiro, and E.~Radu, ``{Effective stability against
  superradiance of Kerr black holes with synchronised hair},'' {\em Phys. Lett.
  B}, vol.~781, pp.~651--655, 2018.

\bibitem{Zouros:1979iw}
T.~J.~M. Zouros and D.~M. Eardley, ``{Instabilities of massive scalar
  perturbations of a rotating black hole},'' {\em Annals Phys.}, vol.~118,
  pp.~139--155, 1979.

\bibitem{Detweiler:1980uk}
S.~L. Detweiler, ``{Klein-Gordon equation and rotating black holes},'' {\em
  Phys. Rev. D}, vol.~22, pp.~2323--2326, 1980.

\bibitem{Suarez:2013iw}
A.~Su\'arez, V.~H. Robles, and T.~Matos, ``{A Review on the Scalar
  Field/Bose-Einstein Condensate Dark Matter Model},'' {\em Astrophys. Space
  Sci. Proc.}, vol.~38, pp.~107--142, 2014.

\bibitem{Hui:2016ltb}
L.~Hui, J.~P. Ostriker, S.~Tremaine, and E.~Witten, ``{Ultralight scalars as
  cosmological dark matter},'' {\em Phys. Rev. D}, vol.~95, no.~4, p.~043541,
  2017.

\bibitem{Arvanitaki:2009fg}
A.~Arvanitaki, S.~Dimopoulos, S.~Dubovsky, N.~Kaloper, and J.~March-Russell,
  ``{String Axiverse},'' {\em Phys. Rev. D}, vol.~81, p.~123530, 2010.

\bibitem{Freitas:2021cfi}
F.~F. Freitas, C.~A.~R. Herdeiro, A.~P. Morais, A.~Onofre, R.~Pasechnik,
  E.~Radu, N.~Sanchis-Gual, and R.~Santos, ``{Ultralight bosons for strong
  gravity applications from simple Standard Model extensions},'' 7 2021.

\bibitem{Hawking:1971tu}
S.~W. Hawking, ``{Gravitational radiation from colliding black holes},'' {\em
  Phys. Rev. Lett.}, vol.~26, pp.~1344--1346, 1971.

\bibitem{Herdeiro:2015tia}
C.~A.~R. Herdeiro, E.~Radu, and H.~R\'unarsson, ``{Kerr black holes with
  self-interacting scalar hair: hairier but not heavier},'' {\em Phys. Rev. D},
  vol.~92, no.~8, p.~084059, 2015.

\bibitem{Wang:2018xhw}
Y.-Q. Wang, Y.-X. Liu, and S.-W. Wei, ``{Excited Kerr black holes with scalar
  hair},'' {\em Phys. Rev. D}, vol.~99, no.~6, p.~064036, 2019.

\bibitem{Delgado:2019prc}
J.~F.~M. Delgado, C.~A.~R. Herdeiro, and E.~Radu, ``{Kerr black holes with
  synchronised scalar hair and higher azimuthal harmonic index},'' {\em Phys.
  Lett. B}, vol.~792, pp.~436--444, 2019.

\bibitem{Herdeiro:2015gia}
C.~Herdeiro and E.~Radu, ``{Construction and physical properties of Kerr black
  holes with scalar hair},'' {\em Class. Quant. Grav.}, vol.~32, no.~14,
  p.~144001, 2015.

\bibitem{Delgado:2016zxv}
J.~F.~M. Delgado, C.~A.~R. Herdeiro, and E.~Radu, ``{Violations of the Kerr and
  Reissner-Nordstr\"om bounds: Horizon versus asymptotic quantities},'' {\em
  Phys. Rev. D}, vol.~94, no.~2, p.~024006, 2016.

\bibitem{Herdeiro:2015moa}
C.~A.~R. Herdeiro and E.~Radu, ``{How fast can a black hole rotate?},'' {\em
  Int. J. Mod. Phys. D}, vol.~24, no.~12, p.~1544022, 2015.

\bibitem{Santos:2020pmh}
N.~M. Santos, C.~L. Benone, L.~C.~B. Crispino, C.~A.~R. Herdeiro, and E.~Radu,
  ``{Black holes with synchronised Proca hair: linear clouds and fundamental
  non-linear solutions},'' {\em JHEP}, vol.~07, p.~010, 2020.

\bibitem{Hod:2012px}
S.~Hod, ``{Stationary Scalar Clouds Around Rotating Black Holes},'' {\em Phys.
  Rev. D}, vol.~86, p.~104026, 2012.
\newblock [Erratum: Phys.Rev.D 86, 129902 (2012)].

\bibitem{Sanchis-Gual:2020mzb}
N.~Sanchis-Gual, M.~Zilh\~ao, C.~Herdeiro, F.~Di~Giovanni, J.~A. Font, and
  E.~Radu, ``{Synchronized gravitational atoms from mergers of bosonic
  stars},'' {\em Phys. Rev. D}, vol.~102, no.~10, p.~101504, 2020.

\bibitem{Cunha:2019ikd}
P.~V.~P. Cunha, C.~A.~R. Herdeiro, and E.~Radu, ``{EHT constraint on the
  ultralight scalar hair of the M87 supermassive black hole},'' {\em Universe},
  vol.~5, no.~12, p.~220, 2019.

\bibitem{Seidel:1993zk}
E.~Seidel and W.-M. Suen, ``{Formation of solitonic stars through gravitational
  cooling},'' {\em Phys. Rev. Lett.}, vol.~72, pp.~2516--2519, 1994.

\bibitem{Brito:2014wla}
R.~Brito, V.~Cardoso, and P.~Pani, ``{Black holes as particle detectors:
  evolution of superradiant instabilities},'' {\em Class. Quant. Grav.},
  vol.~32, no.~13, p.~134001, 2015.

\end{thebibliography}

\end{document}